# Constrained Online Recursive Source Separation Framework for Real-time Electrophysiological Signal Processing

Yao Li [a], Xu Zhang [a,] *, Haowen Zhao [a], Yunfei Liu [a], Xiang Chen [a], Lai Jiang [b, c]

[a] School of Microelectronics at University of Science and Technology of China, Hefei, Anhui, China
[b] Division of Life Sciences and Medicine at University of Science and Technology of China, Hefei, Anhui, China
[c] Department of Obstetrics and Gynecology at the First Affiliated Hospital of University of Science and Technology of China (Anhui Provincial Hospital), Hefei, Anhui, China

Abstract: **Background and Objective:** Processing electrophysiological signals often requires blind source separation (BSS) due to the nature of mixing source signals. However, its complex computational demands make real-time BSS challenging. The objective of this work is to develop an advanced real-time BSS method suitable for processing electrophysiological signals. **Methods:** In this paper, a novel BSS framework termed constrained online recursive source separation (CORSS) was proposed. In the framework, a stepwise recursive unmixing matrix learning rule was adopted to enable real-time updates with minimal computational cost. Moreover, by incorporating prior information of target signals to optimize the cost function, the framework algorithm was more likely to converge to the target sources. To validate its performance, the proposed framework was applied to both downstream tasks, namely real-time surface electromyogram (sEMG) decomposition and real-time respiratory intent monitoring based on diaphragmatic electromyogram (sEMGdi) extraction. **Results:** The proposed method achieved a matching rate of 96.00 ± 2.04% for the sEMG decomposition task and 98.12 ± 1.21% for the sEMGdi extraction task, exhibiting superior performance over other comparison methods ($p < 0.05$). Our method also exhibited minimal time delay during computation, with only 12.5 ms delay when the block size was 0.1s, demonstrating its good capabilities in online processing. **Conclusions:** The proposed method was demonstrated to enable real-time BSS with both improved separation performance and low computational latency. It is of substantial importance for real-time electrophysiological signal processing and applications towards advanced neural-machine interaction and clinical monitoring.

---

*Corresponding author.
E-mail addresses:
yaolee@mail.ustc.edu.cn (Yao Li), xuzhang90@ustc.edu.cn (Xu Zhang)

## 1. Introduction

Electrophysiological signal generated by neuronal discharge reflecting the neural events of human activity [1]. By processing electrophysiological signals collected from different anatomical sites, we can extract precise neural commands from the central nervous system for various human-machine interaction tasks [2], for example, using EEG for emotion recognition [3] and using EMG for myoelectric control [4]. Considering that many tasks require high real-time responsiveness, there are substantial demands for robust online electrophysiological signal processing.

Unlike existing applications that extract simple features from signals in real-time [5], more sophisticated downstream tasks such as signal decomposition and electrophysiological signal denoising necessitate blind source separation (BSS) techniques like independent component analysis (ICA) [6], non-negative matrix factorization (NMF) [7], independent vector analysis (IVA) [8]. The signal model of BSS assumes the multi-channel signal under processing as a mixture of weakly dependent latent non-Gaussian sources [9]. Prior works [10-12] has used BSS method and its variants to achieve significant advancements in electrophysiological signal processing. Among them, the most common objects are electromyography (EMG) and electroencephalography (EEG) signals. However, most approaches are only suited for offline applications [13], for instance, commonly used BSS algorithms such as Infomax ICA and NMF usually have large computational resource consumption, therefore unsuitable for scenarios requiring real-time BSS tasks [14]. Due to the similar nature of these electrophysiological signals, they face the similar challenges when performing BSS operations, that is the difficulty of executing algorithms in real-time. Particularly, non-invasive methods for collecting electrophysiological signals have significant advantages [15], however, the data quality obtained using non-invasive method at the skin surface is generally inferior to that from invasive methods [16], thus making the processing more challenging.

Several researchers have explored methods for performing online BSS. Zhao et al. [13, 17, 18] introduced a series of online BSS methods specifically designed for precise surface EMG (sEMG) signal decomposition. Their approach involves offline computation and adaptive updating of the unmixing matrix (also known as demixing matrix or separation matrix) for online use, thereby achieving stable online performance. While this dual-threaded method has demonstrated good accuracy, its computational demands during offline steps are comparable to conventional ICA methods, indicating no fundamental change in unmixing matrix learning rules. Compared to this kind of method that use unmixing matrix from offline step, another kind of approach is to use methods that changing the unmixing matrix learning rule to achieve real-time BSS by dynamically updating the unmixing matrix using optimized learning rules. These modifications in learning rules are mainly divided into two types: One is based on least mean squares (LMS) and the other type is based on recursive least squares (RLS) [19]. These algorithms with novel learning rules can be used to deal with spatiotemporal nonstationary data commonly encountered in real-world applications [20, 21]. However, the LMS-based algorithms, employing stochastic gradient descent, are computationally straightforward but require careful selection of learning rates for stable convergence [22]. In contrast, RLS-based algorithms accumulate past data exponentially with Sherman-Morrison matrix inversion for faster convergence and improved tracking capabilities, albeit at the cost of increased computational complexity [23, 24]. AKHTAR et al. [25] proposed an online recursive ICA (ORICA) based on an RLS-type recursive rule, which achieves fast convergence and low computational complexity for EEG signal processing. Despite its wild utility in many applications, like other BSS methods, the ORICA algorithm just performs well in extracting signals of interest in a “simplistic” case [26]. When facing the complex mixture of unknown source signals that typically found in real-world applications, it often suffers from incomplete separation among source signals due to the lack of prior knowledge about the target-relevant signals, thereby leading to degraded performance.

Some researchers find that incorporating more assumptions and prior information can avoid local minima and increases the quality of the separation [27]. Stone et al. proposed in [28] that optimal signal separation using Infomax ICA occurred when the contrast function closely matches the cumulative distribution function (CDF) of the target source to be extracted. Building upon this principle, a constrained ICA (cICA) approach that aligns with the probability CDF of electrocardiogram (ECG) signals was designed for real-time cardiac artifact rejection in [29]. Nonetheless, the authors noted that the computation of the unmixing matrix in this method remains overly complex, indicating a need for faster estimation methods.

Given the challenges highlighted in the previous works, this study presents a novel constrained online recursive source separation (CORSS) framework for real-time electrophysiological signal processing. With the recursive learning rule as well as prior knowledge of the target signal to optimize the cost function, the proposed framework facilitates the real-time extraction of precise source signals from surface electrophysiological recordings. This framework was validated through two tasks: real-time sEMG decomposition and real-time extraction of surface diaphragmatic electromyogram (sEMGdi) signals, both utilizing collected experiment data. The proposed framework represents an accurate and robust solution in real-time BSS suitable for electrophysiological signal processing, offering significant implications for real-time myoelectric control, clinical intelligent monitoring, as well as providing valuable insights into building human-machine interfaces.

## 2. Related Works

### 2.1 Blind Source Separation

In BSS tasks, the objective is to separate or recover a set of mixed sources from a set of measured mixtures, which can be expressed as:

$$x(n) = As(n) + v(n) \tag{1}$$

where $x(n) = [x_1(n), x_2(n), \dots, x_M(n)]^T$ is the mixture signal vector with the observed mixture $x_i(n), i = 1,2,\dots,M$, $s(n) = [s_1(n), s_2(n), \dots, s_N(n)]^T$ is the source signal vector with $s_i(n), i = 1,2,\dots,N$ be the scalar source signal vector for a mixing matrix $A$, $v(n) = [v_1(n), v_2(n), \ldots, v_M(n)]^T$ is the vector for uncorrelated additive-random noise, and $n$ is the discrete-time index. The goal is to process $x(n)$ by an adjustable unmixing matrix $W(n)$ such that

$$y(n) = W(n)x(n) \tag{2}$$

contains estimated signals $y(n)$ (also known as component matrix) of source signals $s(n)$. Since the original BSS algorithm such as ICA are typically run in a batch mode [30] in order to keep stochasticity of the empirical gradient low, they are not quite suitable for online implementation in real-time settings.

### 2.2 Online Progressive FastICA Peel-off (PFP)

Chen et al. [31] addressed the issue of FastICA's inability to separate weak source signals in the field of sEMG decomposition by developing an offline progressive FastICA peel-off (PFP) framework. It used constrained FastICA that combines the prior information to access more accurate source signals. Based on this, Zhao et al. [18] introduced an adaptive online PFP method, employing dual-thread operation. They perform offline constrained FastICA to get unmixing matrix updates for online execution, enabling real-time decomposition. This method demonstrates that using constrained FastICA combined with prior information can yield more accurate results. However, this dual-thread approach with offline constrained FastICA inherently leads to increased computational load and latency.

### 2.3 Real-time CICA

Breuer et al. [29] proposed a method based on cICA that enables real-time artifact rejection. They initially utilized the demixing matrix from the preceding data segment as the optimal initial matrix, therefore suitable for real-time calculation. Moreover, they used a block-update rule (also known as block update rule) to update the unmixing matrix of the algorithm based on the prior information for ECG extraction, as described in formula (3) of the Infomax ICA algorithm:

$$\begin{gathered}\Delta W = \eta\left(I + (1-2f){y^*}^T\right)W \\ with\ f = \frac{1}{1 + a_0 . e^{-a_1 y^*}}\end{gathered} \tag{3}$$

where $I$ is the identity matrix, $\eta$ is the learning rate, $y$ is the component matrix, $f$ is the cost function which is fitted to the CDF of cardiac signal, $a_0$ and $a_1$ are constant. By adjusting the cost function, they achieve more accurate result in real time artifact rejection. However, since the basic update rule of the unmixing matrix as shown in formula (3), the author method a faster estimation of the unmixing matrix is desired.

### 2.4 ORICA

AKHTAR et al. [25] proposed an BSS algorithm that achieves incremental updates of the unmixing matrix through a recursive procedure. This method achieves results comparable to those offline ICA with faster convergence. They dynamically update the unmixing matrix in real-time and perform online pre-whitening to achieve single-threaded, convenient real-time separation of source signals, according to formula (4):

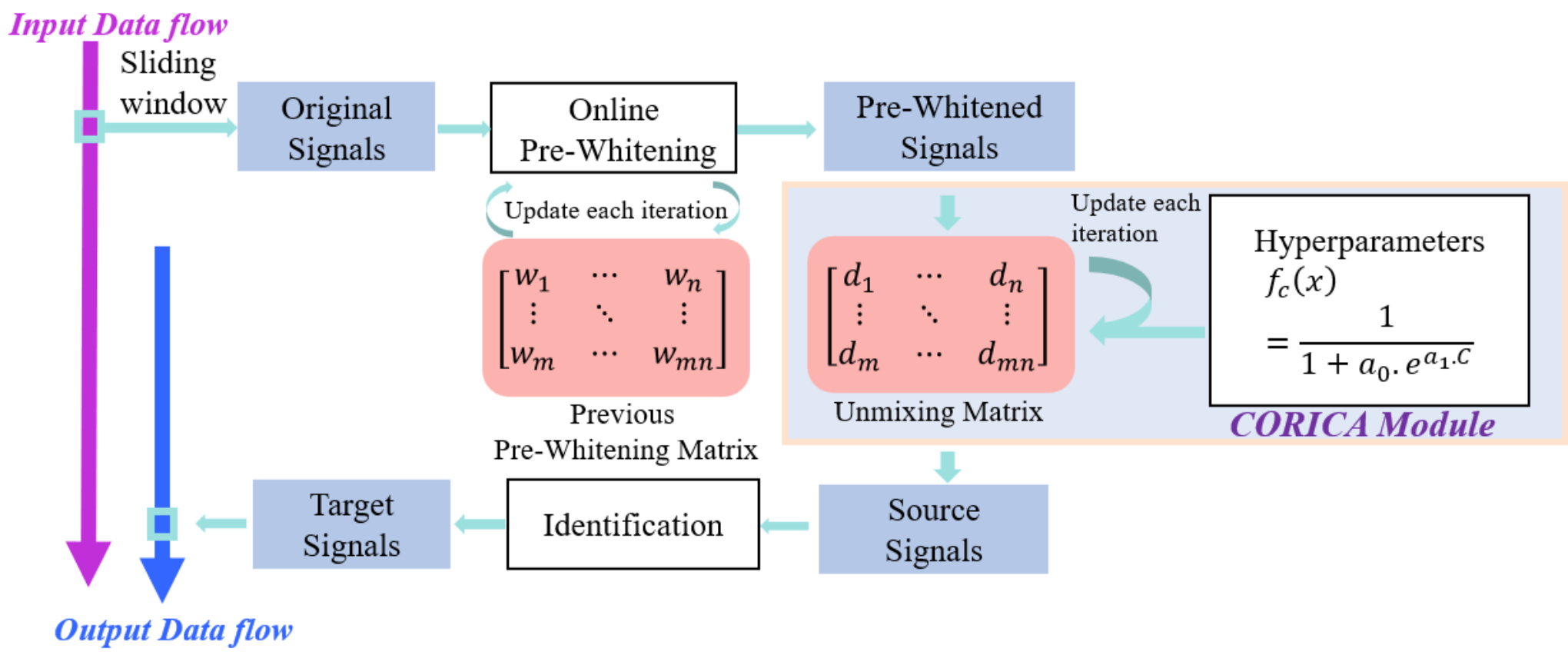

Figure 1. Flowchart of the CORSS framework

$$W_{n+1} = W_n + \frac{\lambda_n}{1-\lambda_n}\left[I - \frac{y_n.f^T(y_n)}{1+\lambda_n(f^T(y_n).y_n-1)}\right]W_n \tag{4}$$

where $W_n$ is the unmixing matrix, $y_n = W_n x_n$ is the component matrix, $\lambda_n$ is a forgetting factor, $f(\cdot)$ is the nonlinear function. With this block-update rule, changes in source patterns and non-stationarity can be addressed, since the ICA model is adaptively updated using the natural gradient of the Infomax ICA rule while the forgetting factor of the algorithm controls the adaptivity to new data and stationarity changes, achieving single-threaded, convenient real-time separation of source signals [32]. However, like most gradient descent algorithms, this approach separates signals at a preliminary level and does not further optimize the algorithm based on the characteristics of the desired signals [26]. Its performance is expected to be unsatisfactory facing complex real-world signals.

## 3. Methods

### 3.1 The CORSS Framework

The flowchart of the proposed CORSS framework is illustrated in Figure 1. To enable real-time data processing, a sliding non-overlapping window was applied to divide the input data flow into consecutive blocks, with the window length, *i.e.* the block size, denoted as *B*. These data blocks served as basic units and raw input fed into the online pre-whitening module for updating the whitening matrix. Following this, the pre-whitened data is directed to the constrained online recursive ICA (CORICA) module, where, based on specific cost function from characteristics of the desired signal, the updated demixing matrix obtained from the previous demixing matrix, according to specific update rules, is used to separate sources that better align with the current moment. After that, an identification process is performed based on the features required for the signal, facilitating real-time acquisition of the desired signal sources.

#### 3.1.1 Online Recursive Pre-Whitening

Most BSS methods require pre-whitening the data, which reduces the number of independent parameters and improves convergence. In this study, an online recursive pre-whitening method [19] is employed for real-time data pre-whitening, as shown in formula (5):

$$M_{n+1} = M_n + \frac{\lambda_n}{1-\lambda_n}\left[I - \frac{v_n \cdot v_n^T}{1+\lambda_n(v_n^T \cdot v_n - 1)}\right] M_n \tag{5}$$

where, $n$ is the number of iterations, $M_n$ is the whitening matrix, $v_n = M_n x_n$ is the pre-whitened data, $\lambda_n$ is the forgetting factor, $I$ is the identity matrix. Through online recursive pre-whitening, we obtained signals that are more suitable for BSS, which are then fed into the next step, the CORICA module.

### 3.1.2 Constrained Online Recursive ICA

Traditional block-update rules for Infomax-based ICA algorithms are known to be computationally expensive and time-consuming [33], and often lack prior information about target signals. Therefore, in this study, we adopt a multiple measurement vector approach [33] and implement updates on short blocks of samples, with prior information about the target signal which correspond to the natural-gradient version of Infomax [34]:

$$W_{n+L} \approx \left(\prod_{l=n}^{n+L-1} \frac{1}{1-\lambda_l}\right) \cdot \left[I - \sum_{l=n}^{n+L-1} \frac{y_l \cdot f^T(y_l)}{\frac{1-\lambda_l}{\lambda_l} + f^T(y_l).\, y_l}\right] W_n$$

$$with \;\; f^T = 1 - \frac{2}{1+a_0 \cdot e^{-a_1 \cdot y_l}} \tag{6}$$

where $W_n$ denotes the previous unmixing matrix and $W_{n+L}$ denotes the updated unmixing matrix, $L$ represents the number of samples in each data block, equivalent to the block size $B$ (e.g., 0.2s) multiplied by the sampling rate (e.g., 1 kHz), $\lambda_n$ is the forgetting factor, $y_l = W_n v_l$ is the component matrix, $v_l$ is the pre-whitened data in the current block, $f(\cdot)$ represents the nonlinear cost function. We used this simple update rule so as to modify the cost function, that is, to determine a set of specific hyper-parameters according to the specific characteristics of the signals, tailoring it to better suit the target signal sources.

### 3.1.3 The Choice of Cost Function

By fitting a function that closely resembles the target CDF as the cost function, prior information about the signal can be incorporated into the computation process [29], yielding results that approach the ideal signal more closely. For specific downstream tasks, modifying the cost function to better fit the target signal facilitates algorithm convergence. Electrophysiological signals often exhibit a pulse-like nature, such as common ECG signals and motor unit action potential trains (MUAPTs) that constitute EMG signals. As illustrated in Figure 2, unlike typical sigmoid or tanh functions for ICA [25], the distribution function of pulsatile electrophysiological signals is approximated based on each pulse cycle by setting a set of hyperparameter.

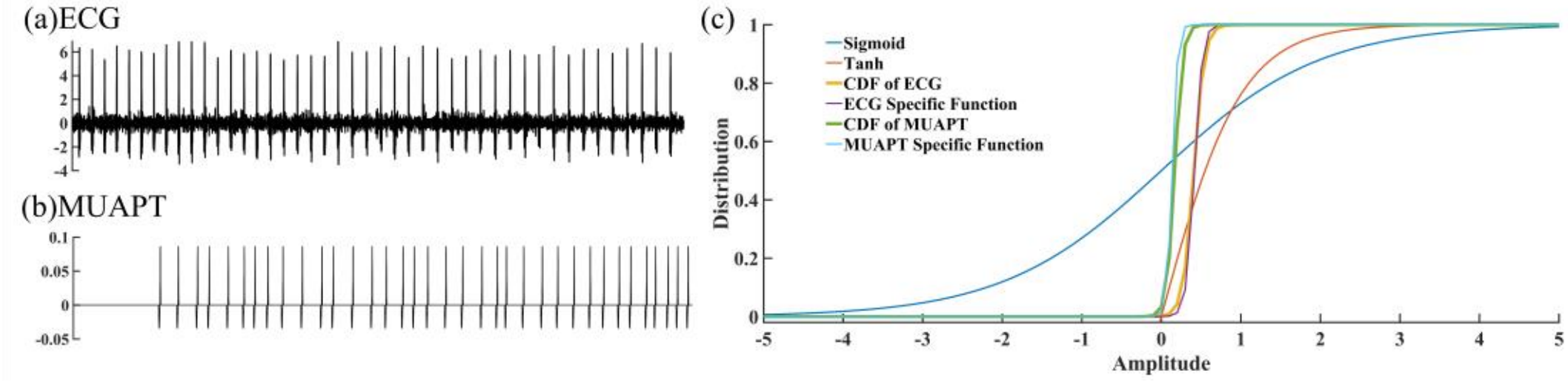

Figure 2. Examples of ECG signal (a) and MUAPT (b), with their CDFs (c) along with common Sigmoid function, tanh function and specific cost functions corresponding to their CDFs, respectively.

### 3.2 Experiment and Data Description

The proposed approach was tested on the following two distinct real-time biomedical signal processing tasks. We evaluated the online processing effectiveness of the proposed method for these tasks to demonstrate its efficacy in real-time electrophysiological signal processing.

#### 3.2.1 Experiments for Real-time sEMG Decomposition Task

As sEMG represents the algebraic summation of MUAPT from a number of active motor units within the electrode's recording range, sEMG decomposition is utilized to decoding MU discharges spike trains and MU action potential trains (MUAPTs) from sEMG signals. The decomposition was also performed using the offline PFP method [31], which has been previously validated with dual-source verification, to extract MUs as the reference.

The sEMG data are collected and pre-processed by the following step: Eight subjects (age: 26.30 ± 1.89 years) without any known muscular injuries or neuromuscular disorders participated in this study. The experimental protocol was approved by the Ethics Review Committee of University of Science and Technology of China (USTC, Hefei, Anhui, China, under Application No. 2022-N(H)-163). Prior to any experimental procedures, all subjects provided informed and signed consent. SEMG signals were recorded from the abductor pollicis brevis (APB) muscle in the dominant hand of each subject. The recording electrode array was a flexible multi-channel setup arranged in an 8 rows × 8 columns configuration. Each electrode probe had a diameter of 2mm with an inter-electrode distance of 4mm between consecutive electrodes. This study employed a reliable, homemade, multi-channel sEMG data recording system, as previously validated in our research [35]. Prior to electrode placement, the APB muscle of the test hand was prepared with medical alcohol. Subsequently, the recording electrode array was fully positioned on the APB muscle. In each data recording trial, subjects were instructed to perform an isometric contraction of the thumb, gradually increasing the muscle force from 0 to the max level in 2 seconds and maintaining it for 3 seconds.

Before proceeding with sEMG data processing, several preprocessing steps were undertaken to reduce noise contamination. Initially, we inspected all the channels of the recorded sEMG signals, and a few channels (typically two to five channels) with a low signal-to-noise ratio were discarded. The sEMG signals within the remaining channels were then passed through a 20–500 Hz Butterworth band-pass filter to eliminate potential low-frequency or high-frequency interference. Power line interference was also removed using a 50 Hz second-order notch filter. The final data input format is $64 \times n$, where $n$ represents the number of sample points (also known as the time length). From a single subject, 5 to 10 signal segments were recorded to form one set of data, with each segment lasting between 30 s to 2 min. Therefore, 8 sets of data contained 57 segments, lasting 2120 s in total.

Since the goal of this task is to extract MU source signals, we set $a_0$ in the cost function to 3 and $a_1$ to -25, which aligns with the PDF of MU according to the design of our framework.

### 3.2.2 Experiments for Real-time sEMGdi Extraction Task

Using sEMGdi for non-invasive respiratory intent monitoring provides direct reflection of neural commands, and is considered state-of-the-art in clinical respiratory monitoring technology [36]. However, surface-collected sEMGdi signals typically come with significant ECG artifacts and noise. In this experiment, we apply the proposed method to separate respiratory signals from 8 sets of actual recordings.

The sEMGdi data are collected and pre-processed by the following step: eight health subjects (mean age: 26.30 ± 1.89 years) participated in the study. The experimental protocol was approved by the Biomedical Ethics Review Board of USTC (Hefei, Anhui, China, under Application No. 2024KY188). Prior to commencing the experiment, informed consent, duly signed by the subjects was obtained.

Multi-channel sEMGdi signals were recorded from the skin overlying the right costal arch of each participant, specifically between the 8th and 10th rib cartilages [37]. A flexible 32-channel monopolar electrode array, arranged in a 4 × 8 grid, was utilized for the recordings. Each electrode had a diameter of 2 mm, with an inter-electrode distance of 4 mm. Data acquisition employed a multi-channel surface electromyogram recording system (FlexMatrix Inc., Shanghai, China), featuring a two-stage amplifier with a total gain of 60 dB, band-pass filtered at 1-500 Hz per channel, and an analog-to-digital converter (RHD 2132, Intan Technologies) sampling at 1 kHz.

Following skin preparation using medical alcohol, the electrode array was positioned over the right costal arch. An Ag-AgCl-based self-adhesive surface electrode served as the reference, placed superior to the tip of the xiphoid process, as depicted in Figure 1(b). During each trial, subject was instructed to breathe at various intensities and rates. The recorded raw sEMGdi data were generally pre-processed to reduce the noise contamination. A Butterworth bandpass filter set at 20-150 Hz was applied to eliminate the potential low-frequency motion artifacts and high-frequency environmental interferences. Subsequently, a set of notch filters were utilized to remove power line interference as well as its harmonics. The final data input format is $32 \times n$, where $n$ represents the number of sample points. From each subject, 4 to 6 signal segments were collected to form a set of recordings, with each segment lasting between 30 s to 2 min. Therefore, 8 sets of recordings contained 42 data segments lasting 2520 s in total.

Since the goal of this task is to separate EMGdi and ECG source signals, we set $a_0$ in the cost function to 0.5 and $a_1$ to -10, which aligns with the PDF of ECG signals for optimal performance.

## 3.3 Performance evaluation

In this study, both tasks involved the sEMG recordings at the same sampling rate of 1 kHz, the block size was optimized after some pretests and constantly set to 0.2 s. For the sEMG decomposition task, in order to assess decomposition performance, the decomposition matching rate (MR) is calculated as follows:

$$MR = \frac{2N_{com}}{N_{online}+N_{ref}} \tag{7}$$

where $N_{online}$ is the number of spikes of each MU spike train from the outcome of online decomposition, $N_{ref}$ denotes the number of the reference spike acquired by offline PFP, and $N_{com}$ denotes the number of common spikes appearing in both online result and reference.

For sEMGdi extraction task, we also use the MR described in Equation 7 to assess respiratory triggering performance. Here, $N_{ref}$ denotes the number of reference breathing triggers, $N_{online}$ denotes the number of correctly identified breathing triggers, when the envelope of the extracted EMGdi signal rises to 30% of its peak value (by calculating the average peak value before), we consider this as triggering a breath. If the timing of this trigger detected by the online method differs from the reference trigger by within 50ms, it is considered a correctly identified trigger time. The sEMGdi extracted using an regular offline method mentioned in [38] served as reference to assess the accuracy of real-time respiratory triggering.

Furthermore, since the task of sEMG decomposition does not involve waveform considerations, we conducted more quantitative analyses on the effectiveness of the methods based on recovered signals in the sEMGdi extraction task. We employ the Root Mean Square Error (RMSE) and correlation (Corr) between the processed signal envelope and the reference envelope to quantify the extent of signal reconstruction:

$$RMSE = \sqrt{\frac{\sum_{i=1}^{n}[EMGdi - \widehat{EMGdi}]^2}{n}} \times 100\% \tag{8}$$

where $n$ denotes the number of sample points, $EMGdi$ denotes the reference signal and $\widehat{EMGdi}$ denotes the estimated signal:

$$CORR = \frac{\sum_{i=1}^{n}(\widehat{EMGdi} - \overline{\widehat{EMGdi}})(EMGdi - \overline{EMGdi})}{\sqrt{\sum_{i=1}^{n}(\widehat{EMGdi} - \overline{\widehat{EMGdi}})^2 (EMGdi - \overline{EMGdi})^2}} \tag{9}$$

where $EMGdi$ denotes the reference signal, $\widehat{EMGdi}$ denotes the estimated signal and $\overline{EMGdi}$ denotes the average of reference signal.

Beside the block size optimally set to 0.2 s, we also evaluate the effect of different block sizes on the computational time delay to evaluate real-time usability of the proposed method. Thus, the block size was varied across 0.1s, 0.2s, 0.4s, 0.5s, 1s and 2s. All of the algorithms were implemented on a desktop computer with an Intel Core i5-10400 processor (2.90 GHz) and 16 GB of memory.

Besides the proposed method, two additional comparison approaches were also employed: Online PFP method with a dual-thread update mechanism [39], and the original ORICA method [25] without constraints specifically designed according to the physiological characteristics. Note that compared to the proposed method, the processing flow of the original ORICA method is identical except for the replacement of the CORICA module with the ORICA module. For the sEMGdi extraction task, we additionally used a commonly employed 50-60 Hz band-pass filtering method [40] for clinical EMGdi extraction as a comparison method.

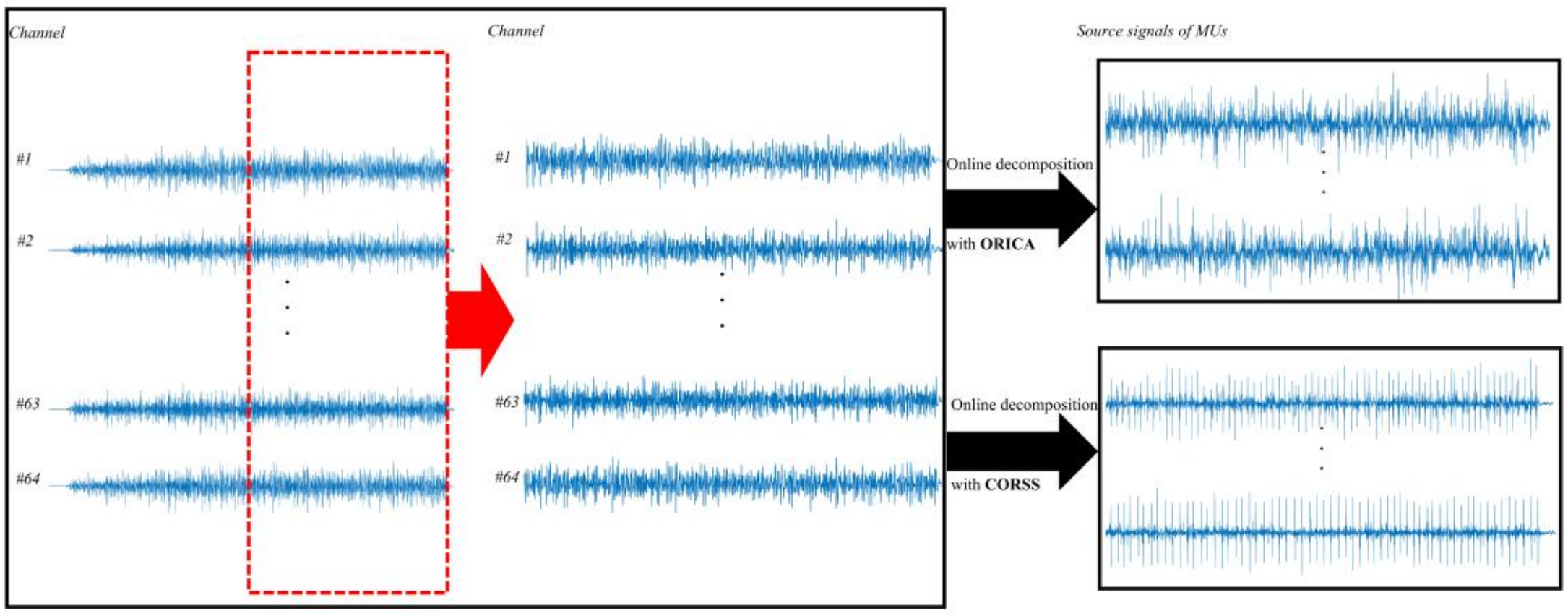

Figure 3. Illustration of sEMG decomposition using different method

### 3.4 Statistical Analysis

In order to examine performance improvement of the proposed method compared with the other comparison methods, a series of one-way repeated-measure ANOVAs were applied to number of MUs and MR, respectively. The homogeneity of variance test and normality test were conducted before each ANOVA. The level of significant difference was set as $p < 0.05$. All statistical analyses were performed by SPSS software (ver. 22.0, SPSS Inc. Chicago, IL, USA)

## 4. Results

### 4.1 The Result of Online SEMG Decomposition

Figure 3 illustrates the preliminary separation performance in terms of MU source signals when processing multi-channel sEMG signals using the ORICA method and the CORSS method, respectively. CORSS clearly reveals distinct sequences of MU source signals, whereas the ORICA method, produces noticeably unusable source signals.

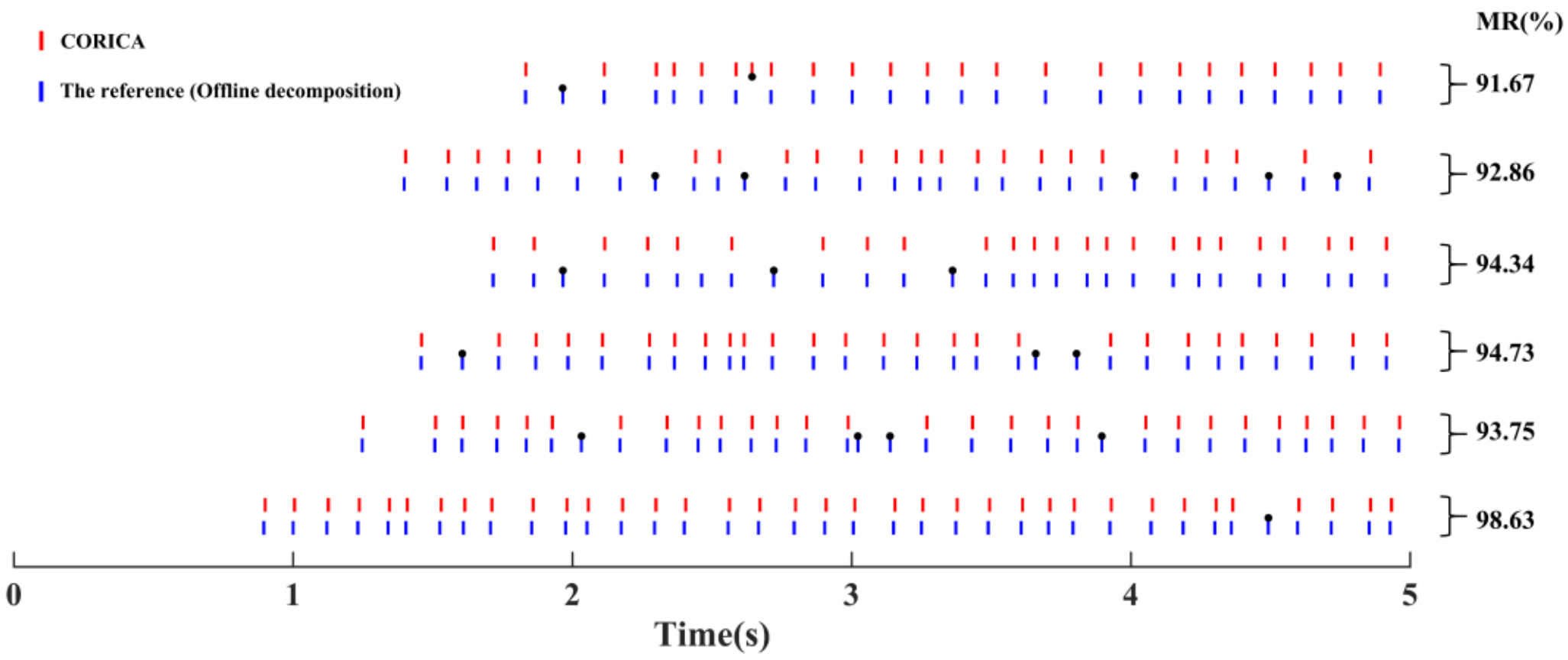

Figure 4. An example of sEMG decomposition using CORSS framework, with the MR of each MU spike train on the right.

Table 1. Summary of results for sEMG decomposition on eight groups of sEMG data.

| No. | Number of motor units | | | | MR (%) | | |
|---|---|---|---|---|---|---|---|
| | Reference | ORICA | Online PFP | CORSS | ORICA | Online PFP | CORSS |
| 1 | 8.01±0.94 | 0.53±0.44 | 6.72±0.35 | 6.67±1.26 | 8.47±3.95 | 97.48±2.94 | 94.25±2.48 |
| 2 | 11.75±1.48 | 0.82±0.76 | 9.20±0.85 | 9.21±0.87 | 10.32±4.21 | 95.63±1.74 | 96.76±1.43 |
| 3 | 7.23±0.82 | 0.64±0.93 | 6.12±0.56 | 6.33±0.90 | 9.95±3.58 | 97.15±1.96 | 98.49±1.90 |
| 4 | 8.42±0.91 | 1.04±0.80 | 5.32±0.32 | 5.09±1.26 | 5.82±8.71 | 95.91±1.41 | 96.07±1.66 |
| 5 | 10.73±1.10 | 1.93±1.39 | 7.75±1.55 | 7.97±0.45 | 13.14±5.18 | 93.48±2.17 | 97.34±2.12 |
| 6 | 9.35±1.23 | 0.73±0.24 | 5.98±1.02 | 4.95±0.91 | 21.66±4.77 | 91.46 ±3.90 | 94.11±3.56 |
| 7 | 12.18±0.72 | 1.04±0.80 | 8.19±0.94 | 9.23±1.14 | 6.05±3.54 | 94.29±3.21 | 94.75±1.85 |
| 8 | 8.50±0.57 | 0.70±0.33 | 4.59±1.21 | 5.42±1.05 | 19.11±7.35 | 98.40±1.26 | 96.25±1.83 |
| Ave. | 9.52±0.95 | 0.93±0.72 | 6.73±1.02 | 6.85±0.87 | 11.82±3.92 | 95.48±1.98 | 96.00±2.04 |

Table 2 Average time delay of three methods with different block size on sEMG decomposition task.

| Block size | Time delay(s) | | |
|---|---|---|---|
| | Online PFP | ORICA | CORSS |
| 100(0.1s) | 0.0125±0.028 | 0.0048±0.012 | 0.0053±0.023 |
| 200(0.2s) | 0.0285±0.073 | 0.0088±0.023 | 0.0083±0.011 |
| 400(0.4s) | 0.0401±0.097 | 0.0146±0.021 | 0.0159±0.038 |
| 500(0.5s) | 0.0394±0.023 | 0.0176±0.018 | 0.0181±0.040 |
| 1000(1s) | 0.0842±0.028 | 0.0187±0.043 | 0.0195±0.028 |
| 2000(2s) | 0.1125±0.018 | 0.0200±0.038 | 0.0213±0.0.21 |

Table 1 further provides quantitative metrics for the proposed method and comparison approaches in sEMG decomposition. The average number of MU spike train identified by the ORICA method are less than one across various levels of reference, as well as an average matching rate of only 11.82% among all sequences. In contrast, the CORSS method and online PFP achieve comparable decomposition results, with matching rates of 96.00% and 95.48% respectively compared to the reference MU spike trains, with statistical significance compared to the ORICA method ($p<0.05$). Figure 4 depicts an example of online decomposition of real sEMG data using the proposed method, showing nearly complete alignment of MU spikes with the reference, albeit with occasional omissions or errors.

Table 2 listed the execution times of online PFP, the ORICA method and proposed CORSS, with different block sizes. It can be observed that for all block sizes, all three methods exhibit a time delay smaller than the block size itself. The time delay of CORSS method varies from 0.0053±0.028s to 0.0125±0.028s with the block size of 100 to 2000, roughly consistent with the time delay of the ORICA method, while the time delay of online PFP is from 0.0125±0.028s to 0.112±0.018s, obviously slower than the other two methods, with statistical significance ($p<0.05$).

**4.2 The result of real-time sEMGdi extraction**

Figure 5 presents an example of 60-second segment of respiratory signals. It is evident that although the comparative method enhances sEMGdi and shows distinct amplitude variations related to respiration, it still exhibits considerable ECG noise, manifested as small spikes in the sEMGdi signal which has a significant impact on its envelope, that can easily lead to incorrect judgment of the triggering moment. In contrast, the proposed method effectively extracts clear sEMGdi with an envelope that obviously conforms to the respiratory patter. This observation is consistently noted across all processed data segments.

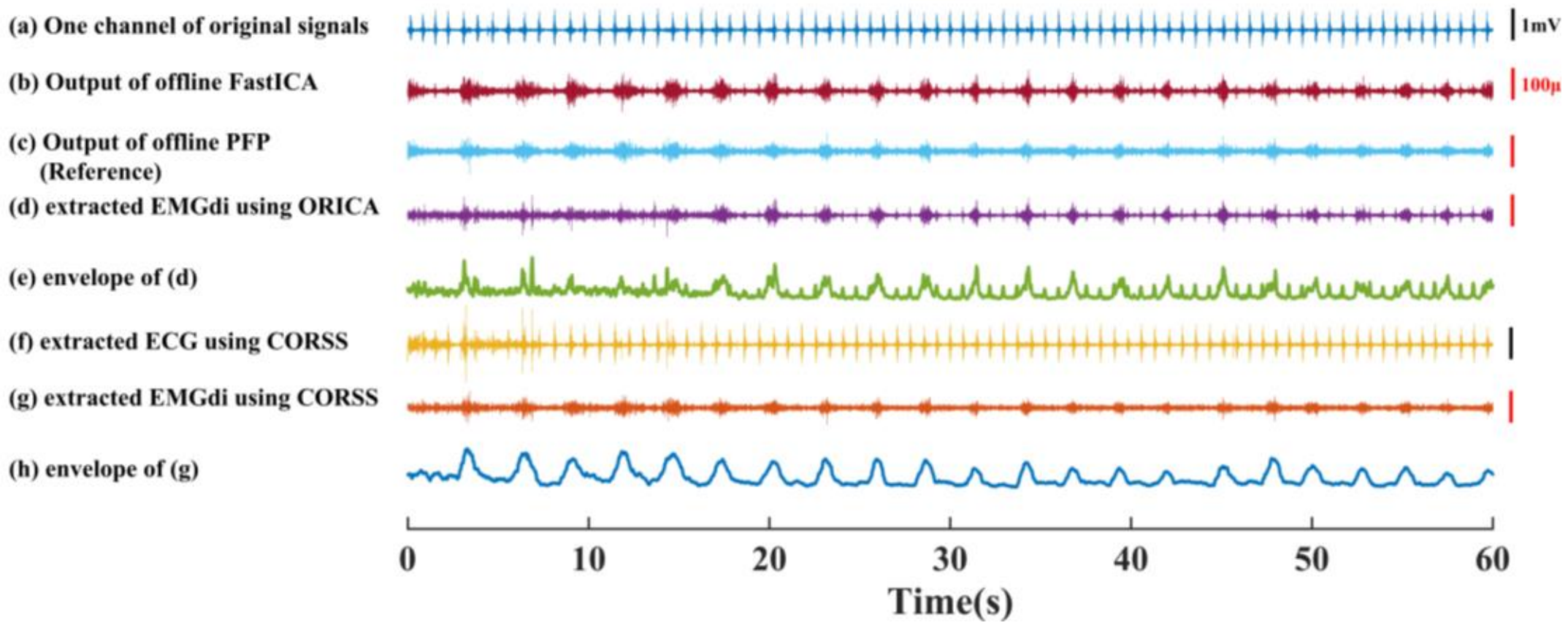

Figure 5. Examples of respiratory-related sEMGdi processed by different method. (a) One channel of original signal. (b) Output of offline FastICA. (c) Output of offline PFP, which is used as reference. (d) Extracted sEMGdi using ORICA. (e) Envelope of extracted EMGdi using ORICA. (f) Extracted ECG using ORICA. (g) Extracted EMGdi using ORICA (h) Envelope of extracted sEMGdi using CORSS.

Table 3 summarizes the quantitative metrics used to describe the performance of sEMGdi extraction, where the RMSE values obtained using the ORICA method and bandpass filter are significantly lower compared to those obtained using online PFP and CORSS, as well as the MR in respiratory trigger recognition, with statistical significance ($p<0.05$). Table 4 shows the execution times (time delay) of four different methods, with different block sizes. It can be seen that, apart from the bandpass filter, all other methods exhibit an increase in time delay with the growing size of a data block. The bandpass filter has the minimum time delay of 0.0039±0.001s even with a block size of 2s, followed by the ORICA and CORSS method, which are on a similar scale. The Online PFP has the maximum time delay.

Table 3. Both RMSE and MR metrics for evaluating sEMGdi extraction performance averaged over eight subjects using four different methods.

| Method | RMSE (%) | MR (%) |
|---|---|---|
| ORICA | 14.72±4.34 | 40.47±12.44 |
| Bandpass Filter | 16.32±2.15 | 31.85±9.54 |
| Online PFP | 8.12±1.21 | 98.95±1.03 |
| CORSS | 6.47±1.80 | 98.12±1.21 |

Table 4 Average time delay of 4 different tasks with different block size on sEMGdi extraction task.

| Block size | Time delay(s) | | | |
|---|---|---|---|---|
| | Online PFP | Bandpass Filter (Single Channel) | ORICA | CORSS |
| 100(0.1s) | 0.0121 ± 0.027 | 0.0037±0.001 | 0.0048±0.012 | 0.0053±0.023 |
| 200(0.2s) | 0.0278 ± 0.069 | 0.0034±0.001 | 0.0088±0.023 | 0.0083±0.011 |
| 400(0.4s) | 0.0390 ± 0.022 | 0.0033±0.002 | 0.0146±0.021 | 0.0159±0.038 |
| 500(0.5s) | 0.0396 ± 0.085 | 0.0034±0.000 | 0.0176±0.018 | 0.0181±0.040 |
| 1000(1s) | 0.0838 ± 0.047 | 0.0026±0.001 | 0.0187±0.043 | 0.0195±0.028 |
| 2000(2s) | 0.1110 ± 0.017 | 0.0039±0.001 | 0.0200±0.038 | 0.0213±0.0.21 |

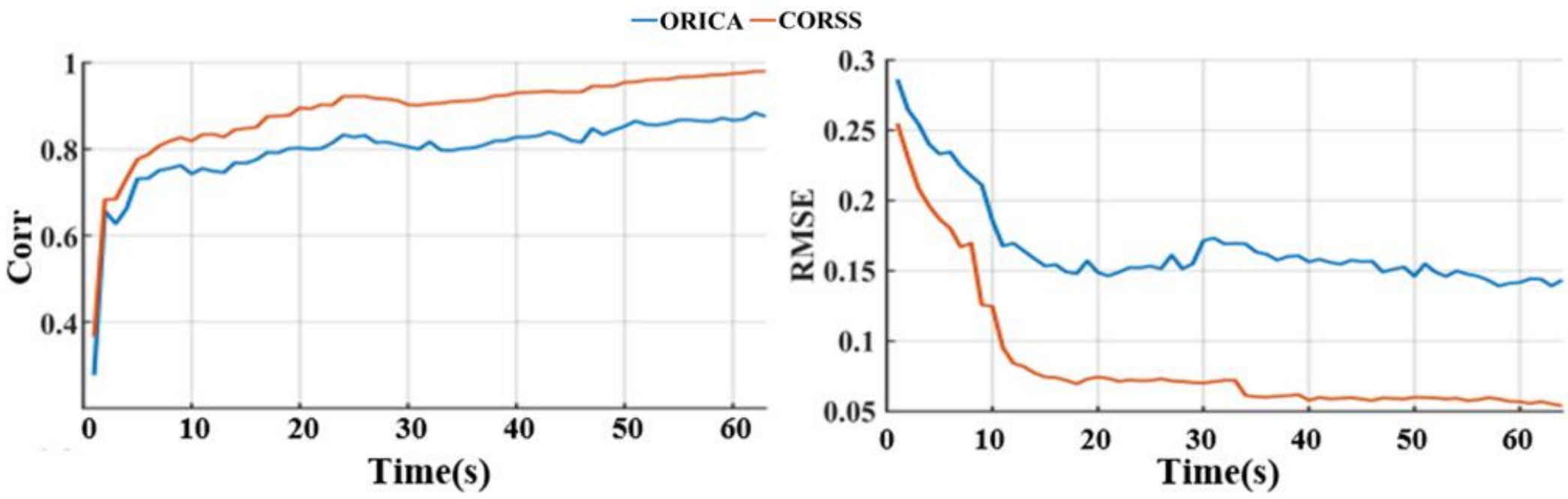

Figure 6. The variations of both Corr (left) and RMSE (right) over time for sEMGdi extraction tasks, averaged over 8 sets of sEMGdi data, using both the ORICA (blue line) and our CORSS method (red line), respectively. The block size was set to 0.2 s (200 samples). Each metric was calculated and averaged just from the data in the past 10 s (50 blocks in total). In the beginning 10 s, the average over the blocks from the initial one to the current one was used.

Figure 6 illustrates changes of both Corr and RMSE metrics with algorithm runtime, using our CORSS method the ORICA method, respectively. It is observed that within the initial 10-20 seconds after algorithm start, both RMSE and Corr are improved obviously. Subsequently, they stabilize without significant change. The signal processed by the CORSS framework achieves a Corr of approximately 0.8 by around 7 seconds, reaching approximately 0.98 thereafter. In contrast, ORICA reaches this level until 20 seconds and stabilizes at approximately 0.87, indicating slower convergence and lower final accuracy compared to CORSS. Similarly, the RMSE obtained by CORSS converges faster than that by ORICA, and the final performance is superior.

## 5. Discussions

Electrophysiological data collected from the human body surface often exhibit overlapping of mixed signals from multiple sources. Current technologies face challenges in real-time separation of any target signal from the source to be concerned. This paper presents a novel real-time BSS framework for processing electrophysiological signals. The proposed method mainly consists of two key techniques:1) a recursive matrix updating rule that enables online BSS with minimal computational resources, and 2) incorporation of prior knowledge associated with source signal characteristics through a cost function, which helps to prevent local convergence and improve separation performance. Such constraint was involved without additional computation approach, further facilitating its real-time implementation. Our method was demonstrated to be effective for both the real-time sEMG decomposition task and the real-time sEMGdi signal extraction tasks.

For the task of sEMG decomposition, the proposed method was observed to extract source signals that are closer to the repetitive firing pattern of MUAPTs than the ORICA method (as shown in Fig. 3), demonstrating its stronger capability of MU source separation. Such source separation improvement can be attributed to prior constraints on pulse-shaped sequences, which is supplemented in the CORICA module in our CORSS framework with respect to the regular ORICA method. On this basis, our CORSS method achieved satisfactory performance in terms of MR of decomposed MUs, comparable or even superior to that of the Online PFP method, with no statistical significance ($p > 0.05$). The later one was reported as the SOTA method intentionally designed for online decomposition of sEMG signals, indicating advanced performance of the proposed method for this task. Notably, the proposed method shows a significant reduction in computational time delay compared to Online PFP, suggesting that it has better computational efficiency and is more suitable for real-time processing applications. This advantage inherits the high computational efficiency from the ORICA method, which yielded the lowest computational delay but unsatisfactory source separation performance. With these regards, the proposed method is proven to juggle both high performance and real-time computational efficiency.

It is noteworthy that using ICA with constraint is a crucial step in processing electrophysiological signals, allowing for the separation of accurate target signal. The method proposed in this paper introduces the CORICA module, which incorporates constraints by using a source specific cost-function and a recursive unmixing matrix that contains the prior information [29]. This allows the ORICA method to gain advantages suitable for processing physiological signals, thereby explaining its superior performance in two tasks. In contrast, while the Online PFP method demonstrates strong performance, its approach to adding constraints is by measuring the similarity between a reference signal and the extracted sources [31], which significantly add high computational cost. This explains the notably higher latency of online PFP observed in this study. Consequently, the proposed method can be viewed both as an advanced performance upgrade for BSS methods with online computation capabilities and as a significant improvement in the applicability of ORICA for complex electrophysiological analysis.

Besides the sEMG decomposition task, the success of the proposed method in sEMGdi extraction further demonstrates its broad applicability. The sEMGdi extraction task, aimed at NAVA ventilator applications, faces higher demands and challenges for real-time computation. Although ORICA is considered to have optimal real-time performance, its separation performance is unsatisfactory because it cannot effectively isolate sEMGdi from strong ECG interference. As a result, the extracted sEMGdi contains some ECG-related fluctuations, which affect the calculation of the sEMGdi envelope. This interference can significantly impact the expression of respiratory intention information that drives the ventilator, introducing application risks [36]. In contrast, the CORSS method can extract clean sEMGdi signals with minimal influence from strong ECG interference, producing a smooth sEMGdi envelope that meets the needs of driving the NAVA ventilator. Although methods using offline FastICA and ORICA differ in computational processes, their final outcomes are largely similar due to their lack of relevant information about the target signal, thus less accurate, as shown in Fig. 5. Note that we use the PDF of ECG as the cost function in this task. Since the two main sources in this task are ECG and sEMGdi, accurately identifying one source also helps to clearly separate others.

When considering precision of extracted signal waveforms, both Corr and RMSE metrics were found to improve towards their optimal direction rapidly within 20 s, as shown in Figure 6. This suggests that our method was able to work effectively after about 20 s of training with and adaptation to the input signals, enabling continuous update of the demixing matrix in real-time. Even with significant signal variations (such as electrode displacement during measurement), the algorithm swiftly maintains to work well after brief updates. Such rapid convergence time is crucial for clinical applications. As mentioned in [41], a common empirical heuristic time for the number of training samples required for separating $N$ stable ICA sources using traditional Infomax-based ICA was $kN^2$, where $k > 25$. Therefore, for our 32-channel signal, the heuristic time required for convergence should be amounted to $32^2 \times 25 = 25600$ samples=25.6 second with a 1000Hz sample rate. Our findings were consistent with this theoretical calculation.

A series of evaluations on these representative tasks demonstrates that the proposed method represents a mechanistically superior approach compared to existing methodologies, which holds significant advantages in real-time processing of electrophysiological signals involving BSS. Its relevance extends to real-time human-machine interaction and clinical monitoring. Given its fast implementation and advanced performance, our method is also expected to show feasibility and usability in other extensive tasks such as clinical ECG monitoring and real-time control of assistive devices.

Currently, a major limitation of this method is its dependence on empirical judgment for adjusting prior parameters. In the future, we can explore methods for automatically adaptation of prior knowledge based on signal characteristics, enabling the framework to intelligently adapt to a broader range of tasks.

## 6. Conclusion

In this study, we propose a constrained online recursive source separation (CORSS) framework for real time electrophysiological signals processing. With a stepwise recursive unmixing matrix learning rule, the algorithm achieves real-time updates with minimal computational overhead. The framework was tested on different downstream tasks with state of the art performance. Our approach demonstrates strong performance across various real-time electrophysiological signals processing tasks, highlighting its significance for applications in real-time human-computer interaction and clinical monitoring.

Acknowledgments

This work was supported by the Anhui Provincial Key Research and Development Program under Grant 2022k07020002.